\documentclass{jltp}

\usepackage{graphicx,epsfig} 

\newcommand{\beq}{\begin{equation}}
\newcommand{\eeq}{\end{equation}}
\newcommand{\bea}{\begin{eqnarray}}
\newcommand{\eea}{\end{eqnarray}}

\newcommand{\veps}{\varepsilon}
\newcommand{\etal}{{\it et al.},}
\newcommand{\nn}{\nonumber}

\title{ Signatures of Superfluidity in Dilute Fermi Gases near a
         Feshbach Resonance }

\author{Aurel Bulgac$^1$ and Yongle Yu$^2$}

\address{$^1$Department of Physics, University of Washington,\\
Seattle, WA 98195-1560, USA\\
$^2$ Division of Mathematical Physics, Lund Institute of Technology,\\
Box 118 SE 221 00 LUND, SWEDEN}

\runninghead{A. Bulgac and Y. Yu}{Signatures of Superfluidity in Dilute 
Fermi Gases}

\begin{document}

\maketitle

\begin{abstract}

We present a brief account of the most salient properties of vortices
in dilute atomic Fermi superfluids near a Feshbach resonance.

\end{abstract}

\section{INTRODUCTION}

Since the creation of the first degenerate Fermi gas by DeMarco and
Jin of $^{40}$K atoms and the optical trapping of $^6$Li atoms by
Thomas' group\cite{jin} the last year and a half has produced an
enormous experimental progress in the study of dilute atomic Fermi
gases. The ability to manipulate the strength of the interaction by
means of the Feshbach resonance opened extraordinary opportunities
both from the experimental and theoretical points of view. There is no
doubt in anybody's mind that dilute atomic Fermi clouds near a
Feshbach resonance should become superfluid, with a pairing gap of the
order of the Fermi energy at sufficiently low temperatures.  The
challenge from the experimental point of view is to realize these
superfluids and especially to demonstrate unambiguously the onset of
superfluidity, or at least the formation of a condensate and be able
to study its properties. A real breakthrough was the creation and the
subsequent study of the expansion of a strongly interacting degenerate
Fermi gas\cite{duke}.  After that experimentalists have been able to
study the formation of extremely weakly bound molecules (which we
shall often refer to as dimers)\cite{molecules}, the decay properties
of ensembles of such dimers\cite{decay}, the BEC of dimers\cite{bec},
a number of features of the BCS to BEC crossover\cite{bcsbec}, the
collective oscillations\cite{oscillations}, the formation of some kind
of condensate, with some still unclear properties\cite{condensation}
and finally the appearance of a gap in the excitation
spectrum\cite{gap}.

Leggett and others have envisioned theoretically such a BCS to BEC
crossover\cite{leggett} and were able to describe qualitatively its
main features. Qualitative features of the BCS dilute atomic Fermi
superfluid have been discussed by a number of authors in recent
years\cite{eddy}. The theoretical description was based essentially on
the weak coupling BCS formalism, which is known to over predict the value
of the gap by a significant factor\cite{gorkov}. The crossover theory
of Leggett and its followers was based on a more or less
straightforward extension of the weak coupling BCS formalism to the
strong coupling regime. In the BEC limit there is an equally
significant correction of this results\cite{amm}. As it was noted by
Bertsch\cite{george}, a dilute Fermi system acquires universal
properties at, what nowadays we call, the Feshbach resonance. The
initial studies of the Bertsch MBX challenge showed that such a
system is stable\cite{baker,heiselberg}. Only relatively recently that was
confirmed both theoretically\cite{carlson,chang,giorgini} and
experimentally\cite{duke}.

The discussion of some general properties of these systems, the
character of the collective oscillations in trapped dilute atomic Fermi
gases near a Feshbach resonance, which we planned to cover as well,
along with the discussion of the properties of atom-dimer mixtures,
see Refs. \onlinecite{bbb}, are skipped here due to space limitations.

\section{Superfluid LDA and the Vortex State}

From the theoretical point of view the challenge is to be able to
predict and describe in a controllable manner the properties of these
systems, which necessitates the development of accurate theoretical
tools. We have shown recently how to extend the density functional
theory \cite{kohn} to superfluid fermion systems, by creating the so
called Superfluid LDA (SLDA)\cite{slda,vortex}.  In the case of SLDA
one needs to know the dependence of the energy density as a function
of both normal and anomalous densities, unlike LDA when only the
dependence on the normal density is sufficient. The rather accurate
calculations of Refs. \onlinecite{carlson,chang} allows us to
construct the energy density functional (EDF) in the case of infinite
homogeneous systems. Using the recent results of Chang {\it et al.} in Ref.
\onlinecite{chang} we decided to re-parameterize this EDF and extend
that parameterization away from the Feshbach resonance. In terms of
the single quasi-particle wave functions, which define the normal and
anomalous densities the EDF of a superfluid system in the SLDA
approach has the form:
\bea
& & \!\!\!\!\!\!
{\cal{E}}_S({\bf r}) n ({\bf r}) =
\frac{\hbar^2}{m}
\left \{
\frac{1}{2}\tau({\bf r})n ({\bf r}) +
\beta\left [\frac{1}{n({\bf r})a^3}\right ] n ({\bf r}) ^{5/3} +
\gamma\left [ \frac{1}{n({\bf r})a^3} \right ]
\frac{|\nu({\bf r})|^2}{n({\bf r})^{1/3}}
\right \},\nn \\
& & n ({\bf r}) =\sum_\alpha |v_\alpha ({\bf r})|^2,\quad
   \tau ({\bf r}) =\sum_\alpha |{\bf \nabla}v_\alpha ({\bf r})|^2, \quad
  \nu ({\bf r}) =\sum_\alpha v_\alpha^* ({\bf r})u_\alpha ({\bf r}), \nn
\eea
where the spin degrees of freedom have been suppressed for the sake of
simplicity. The dimensionless functions $\beta(x)$ and $\gamma(x)$ can
be easily determined using the recent the results of
Ref. \onlinecite{chang}. The SLDA equations have the same formal
structure as the Bogoliubov-de Gennes equations for a system with
density dependent contact interaction. In spite of the formal
resemblance these equations describe such systems exactly, beyond the
meanfield approximation, namely the ground state energy and number
density distribution. 

The local anomalous density $\nu({\bf r})$ is
unfortunately strictly speaking a diverging quantity and its
evaluation requires a well defined regularization procedure. The
evaluation of the pairing field also requires a renormalization
procedure and so does actually the evaluation of the total ground
state energy. The principles of the regularization and renormalization
procedures have been described by us previously, see
Refs. \onlinecite{slda,vortex}. In the region of the Feshbach
resonance, where the pairing gap is of order of the Fermi energy
$\veps_F = \hbar^2(3\pi^2n)^{2/3}/2m$ these procedures have to be
changed somewhat, in order to ensure a better convergence of various
quantities, number density, pairing field, total energy, etc. and
these details will be presented elsewhere\cite{njp}. It suffices to
add that unlike a number of other methods suggested recently in
literature, aimed at dealing with similar divergences, see
Refs. \onlinecite{bruun} and discussion in Ref. \onlinecite{urban},
and which require an active Hilbert space of a size $10^3 \ldots
10^5\; N$, in our approach the size of the active Hilbert space is
typically much less than $10N$, where $N$ is the total number of
particles. 

The form of the EDF presented above is not unique to a
certain extent. In particular one could have considered a density
dependent mass in the kinetic energy term, as is often done in nuclear
physics. Until full many-body calculations of the homogeneous matter
of the kind described in Refs. \onlinecite{carlson,chang,giorgini}
will provide more detailed information about the properties of such
systems, beyond the energy per particle and the pairing gap as a
function of the parameter $1/na^3$, there is no unambiguous way one can
determine whether the effective mass is different or not from the bare
mass. There is no spin-orbit coupling as well and so far there is no
compelling argument to expect its presence. However, one can expect
gradient terms, in particular a dependence of the EDF on ${\bf
\nabla}n({\bf r})$. In principle such terms could be evaluated, but
their role is not expected to be ever dominant, though it could be
significant. Phenomenologically, in nuclear physics it is established
that such terms are quite important. The reason they are important in
nuclear physics is because the radius of the interaction is comparable
with the Fermi wave length. This is definitely not the case of dilute
atomic gases for which $nr_0^3 \ll 1$ always (here $r_0$ is of the order
of at most 100 \AA $\;$ or so, the so called van der Waals length). One can
come up with a similar qualitative argument in favor of an effective
mass close in value to the bare mass, which was implicitly assumed by
us. It is worth noting also that the present EDF is quite distinct
from others suggested recently in literature, see for example
Ref. \onlinecite{perali}, which are typically based on one or another
incarnation of the crossover model due to Leggett\cite{leggett}.

\begin{figure}
\centerline{\includegraphics[height=3.5in]{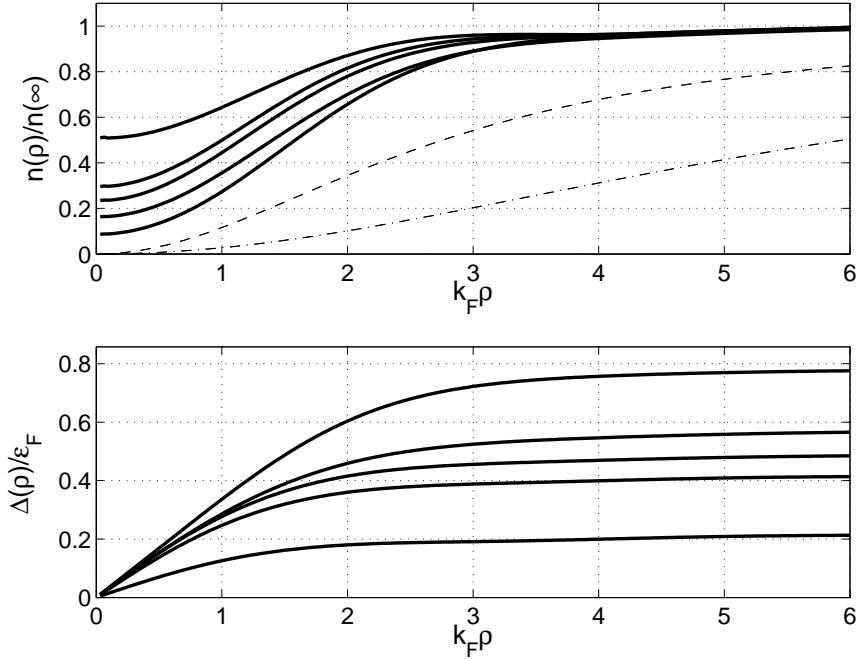}}
%
\caption{The upper panel shows the number density profile around a
  vortex core, while the lower panel shows the profile of the pairing
  field. In the upper panel various curves correspond to
  $1/k_Fa=(-0.5,\; -0.1,\; 0,\; 0.1,\; 0.3)$, from the highest to the
  lowest respectively. The order is reversed in the lower panel. We
  show also the (approximate) density profile of a vortex in a Bose
  dilute atomic gas of the same number density, with a dashed line for
  the case $na^3 =10^{-3}$ and with a dot-dashed line for the case
  $na^3=10^{-5}$.}  
\label{fig:tau2}
\end{figure}

At this point all the elements needed, in order to perform a full
self-consistent calculation of the vortex properties in a dilute atomic
Fermi gas near a Feshbach resonance, are known. Without dwelling into
technical details\cite{vortex,njp} we shall briefly discuss the most
salient features. In Fig. 1 we show the number density profile $n({\bf
\rho})/n(\infty)$ and the profile of the pairing field $\Delta({\bf
\rho})/\veps_F$ around a vortex core, where ${\bf \rho}= (x,y)$ and
the $z$-axis is along the vortex core. At the Feshbach resonance the
asymptotic value of the pairing field is approximately one half the free
Fermi energy of the free gas. The actual Fermi energy at the Feshbach
resonance is $\approx 0.8\veps_F$. As one can see from the lower
panel in Fig. 1 the pairing field has a rather dramatic dependence on
the scattering length\cite{chang}. 

What came as a big surprise
before\cite{vortex} and it is confirmed by the present results, based
on a more accurate EDF, is the unexpected appearance of the prominent
vortex core number density depletion shown in the upper panel of
Fig. 1. For comparison we show there as well the number density
profile of two vortices in dilute Bose gases of the same density and
two different values of the corresponding atomic scattering length. As
one can see the size of the vortex core is only 2-4 times smaller in
the Fermi case than the vortex core in the BEC case. This fact let us
conclude that a direct visualization of vortices should be easily
achievable in the case of atomic Fermi superfluids around the Feshbach
resonance.

\section{Conclusions}

The onset of superfluidity should undoubtedly be demonstrated by
exciting and putting in evidence a superflow, and vortices are just
about the only such modes in which superflow can be seen in such
systems. Moreover, these vortices are also expected to form an
Abrikosov lattice. We presented an analysis, based on a newly
developed extension of LDA to superfluid systems, Superfluid LDA
(SLDA), of the properties of vortices at and near a Feshbach
resonance. Surprisingly, like in the case of Bose superfluids,
vortices in Fermi superfluids near a Feshbach resonance share common
features, which should make them easily detectable. Vortices develop a
pronounced density depletion in the core of a size comparable to the
size of a core in a dilute atomic Bose gas.

\end{document}